\documentclass[onecolumn,11pt,a4paper]{article}

\usepackage{graphicx}
\usepackage{float}

\title{High Density and Non-volatile CRS-based CAM}

\begin{document}

\begin{center}
{\LARGE {Memristor-based Synaptic Networks and Logical Operations Using In-Situ Computing}}
\end{center}

\vskip 1.2em%

\large{
\begin{center}
{Omid Kavehei{\small $~^{1}$}, Said Al-Sarawi{\small $~^{1}$}, Kyoung-Rok Cho{\small $~^{2}$}, Nicolangelo Iannella{\small $~^{1}$}, Sung-Jin Kim{\small $~^{2}$}, } 
{Kamran Eshraghian{\small $~^{2}$}, Derek Abbott{\small $~^{1}$} }
\vspace{1.6mm}\\
\fontsize{10}{10}\selectfont\itshape
$^{1}$\,School of Electrical \& Electronic Engineering, The University of Adelaide, SA 5005, Australia \\
$^{2}$\,College of Electronics \& Information Engineering (WCU Program), Chungbuk National University, South Korea \\
\fontsize{9}{9}\selectfont\ttfamily\upshape
omid@eleceng.adelaide.edu.au
\end{center}
}

{\it Abstract}-We present new computational building blocks based on memristive devices. These blocks, can be used to implement either supervised or unsupervised learning modules. This is achieved using a crosspoint architecture which is an efficient array implementation for nanoscale two-terminal memristive devices. Based on these blocks and an experimentally verified SPICE macromodel for the memristor, we demonstrate that firstly, the Spike-Timing-Dependent Plasticity (STDP) can be implemented by a single memristor device and secondly, a memristor-based competitive Hebbian learning through STDP using a $1\times 1000$ synaptic network. This is achieved by adjusting the memristor's conductance values (weights) as a function of the timing difference between presynaptic and postsynaptic spikes. These implementations have a number of shortcomings due to the memristor's characteristics such as memory decay, highly nonlinear switching behaviour as a function of applied voltage/current, and functional uniformity. These shortcomings can be addressed by utilising a mixed gates that can be used in conjunction with the analogue behaviour for biomimetic computation. The digital implementations in this paper use in-situ computational capability of the memristor.

\section{Introduction}

The classical von Neumann machine suffers from a large sequential (fetch-execute-store cycle) processing overload due to the existence of the data bus between memory and logic. Neuromorphic engineering introduces a more efficient (event driven) implementation but not necessarily low-power. Software techniques are power hungry and there traditionally has been was no low-power hardware device (switch) to provide tighter coupling between memory and logic, as in biological systems. The memristor is an emerging technology that combines (non-volatile) memory and in-situ computational characteristics in one device in the way that promises an entirely new computer architecture.

The mathematical foundation of the memristor, as the fourth fundamental passive element, has been expounded by Leon Chua~\cite{memristor:chua:1971} and later extended to a broader class, known as memristive devices and systems~\cite{memristor:chua:1976}. This broad classification today includes all resistance switching memory devices~\cite{resistance:chua:2011}. Realisation of a solid-state memristor in 2008~\cite{memristor:strukov:2008} has generated a new wave of research in realization of both large memory arrays as well as new thinking in the neuromorphic engineering domain. Memristors (the term {\it memristor} is a portmanteau of {\it memory} and {\it resistor}) are capable of encoding information in two or more stable levels each with relatively long decay times. The decay can be long in human terms (e.g. days and weeks), which is a practical implementation of a non-volatile memory~\cite{memristor:chua:1976,memristor:strukov:2008,fourth:kavehei:2010,instar:snider:2011,learning:hasegawa:2010}. It has also been experimentally proven -in small scale- that these two-terminal memristive devices are able to carry out logic operations~\cite{memristive:borghetti:2010}. Therefore, memristor is a possible option for implementing a tighter coupling between memory and logic technologies. 

There are many memristor-based applications. The obvious application of such a nanometer scale device is in implementing non-volatile, low-power, and dense memory arrays. Owing to the multi-stable state property and the relatively long term decay, memristors are also able to encode synaptic weights~\cite{nanoscale:jo:2010}. Furthermore, several possibilities for neuromorphic engineering domain and learning have been also studied~\cite{memory:pershin:2011,instar:snider:2011, cortical:snider:2008, spike:zamarreno:2011}. In this paper we demonstrate very basic analogue and digital circuits that are implemented in memristor technology.

Contributions that this paper provides can be categorised as follow:
\begin{itemize}
		\item Brief characterisation of memristor for neuromorphic purposes. 
		\item Experimental results demonstrating the multi-stable state of a silver/titanium dioxide/indium tin oxide (Ag/TiO$_2$/ITO).
		\item Demonstration of the use of memristor as a synaptic connection that mimics the Spike-Timing Dependent Plasticity (STDP) rule. 
		\item Show a memristor-based competitive Hebbian learning through STDP. 
		\item Circuit for analogue multiplication and accumulation using fixed weights pattern. 
		\item Experimental show that a sharp switching behaviour in a fabricated Ag/TiO$_2$/ITO and Pt/TiO$_2$/Pt (Pt: Platinum) memristors as well as state decay. Demonstrating a memristive-based analogue computing. 
		\item CRS-based logic gates through material implication and PLA implementations.
\end{itemize}	

Note that memristor and memristive device characteristics, modelling, materials, and underlying physics are not within the scope of this paper. The interested reader can find further details in~\cite{memristor:chua:1971,fourth:kavehei:2010,memory:pershin:2011,analytical:kavehei:2011,fabrication:kavehei:2011} for further details. The simulations carried out in this work using SPICE macro-model implementation of presented model in~\cite{analytical:kavehei:2011}. 

\section{Memristor model} \label{sec:memmodel}

Memristor device characteristics can be defined using a system of two equations,
\begin{eqnarray}
\left\{ \begin{array}{ll}
I=g(w,V)\cdot V\\\label{equ:mem}
\frac{{\rm d}w}{{\rm d}t}=f(w,V)~,\end{array}\right.\\ \nonumber 
\end{eqnarray} 
where $w$ is a physical variable indicating the internal memristor state that in theory is such that $0<w<L$, $L$ is the thickness of a thin-film metal-oxide (memristive) material sandwiched between two metallic electrodes, and $I$ and $V$ represent current and voltage, respectively. The $g(\cdot)$ function represents the memristor's conductance. The state variable can be expressed using a normalised form $x=1-w/L$. In this case, $w\rightarrow 0$ or moving towards higher conductances can be expressed as $x\rightarrow 1$ and $w\rightarrow L$ or moving towards lower conductances can be shown as $x\rightarrow 0$. In this paper, $R_{\rm HRS}$ represents high resistance state and $R_{\rm LRS}$ shows low resistance state. Eq.~(\ref{equ:mem}) shows that the output of the system (here $I$), at a given time, depends on $w$ and $V$. State transition conditions are also explained by the function $f(\cdot)$. To measure this function several time-domain experiments for $I$ and $V$ are required. According to our measurements, a $\sinh(\cdot)$ like behaviour can explain dynamics of the device while an additional term is needed. The $\sinh(\cdot)$ term defines the dependency of velocity, $dw/dt$, to the effective applied electric field that has been described as an ionic crystal behaviour in an external electric field~\cite{electronic:mott:1964}. The additional term highlights the dependency of conductance, $G_{t}$, to the previous conductance, $G_{t-1}$. Intuitively, we use an exponential form function $h(w)$ to define $dw/dt$ as a function of $w$ based on Fig.~3 in~\cite{switching:pickett:2009}. The $h(w)$ function then should be multiplied by the $\sinh(\cdot)$. The conductance behaviour as a function of $w$ is also shown in Fig~2 of~\cite{fabrication:kavehei:2011}. Due to the asymmetric behaviour of $w\rightarrow 0$ and $w\rightarrow L$~\cite{switching:pickett:2009}, we have used two different $h(w)$ definition to address a more accurate switching properties~\cite{switching:pickett:2009, fabrication:kavehei:2011}. 

The state variable equation then can be defined as 
\begin{eqnarray}
\frac{dw}{dt}=h(w)\sum_{i}\upsilon_iV^i+d(w),\label{equ:newvelocity}
\end{eqnarray} 
where $\upsilon_i$ are coefficients for low and high electric fields. The index, $i$, is an positive odd integer so it is the expansion of $\sinh(\cdot)$. This demonstration help to easily extract linear approximation of the memristor model in~\cite{memristor:strukov:2008} and also combine effects of Joule heating and $L-w$ (the effective electric field) in the coefficients~\cite{analytical:kavehei:2011}. The function $d(w)$ represents the decay term which can be weeks, months, or more. The decay term appears to be very similar to synaptic weight update (learning) rule~\cite{instar:snider:2011,cortical:snider:2008}. The first term of Eq.~(\ref{equ:newvelocity}), represents a voltage dependent, highly nonlinear which makes high-speed digital computing possible. This property originated from the fact that resistance modulation inside the metal-oxide occurs via electron-ion interactions. This term creates a significant problem for learning applications in the current form. 

To compensate this problem we have to take advantage of its high nonlinearity. This nonlinear behaviour produces a threshold-like region that voltages below that threshold does not change the conductance. Considering the fact that, memristor's conductance, $G$, can be tuned by a series of voltage pulses with appropriate pulse widths and a voltage around the threshold, obviously, pulse time is the other parameter involved. Applying a voltage around the threshold slightly changes $x$ (or $w$) if it is maintained for a few $\mu$s. It is observed that such voltage cannot change the state if the duration is around a few ns. However, slightly increase in the applied voltage increases the speed by several orders of magnitude, which makes nanosecond (digital) switching possible. Therefore, a series of few $\mu$s pulses with an appropriate pulse shape can be used to mimic learning rule~\cite{spike:zamarreno:2011}.

\section{Analogue Memory and Computing}

\subsection{Muti-stable state memory}

\begin{figure}[htpb] \centering
  \includegraphics[width=0.6\textwidth]{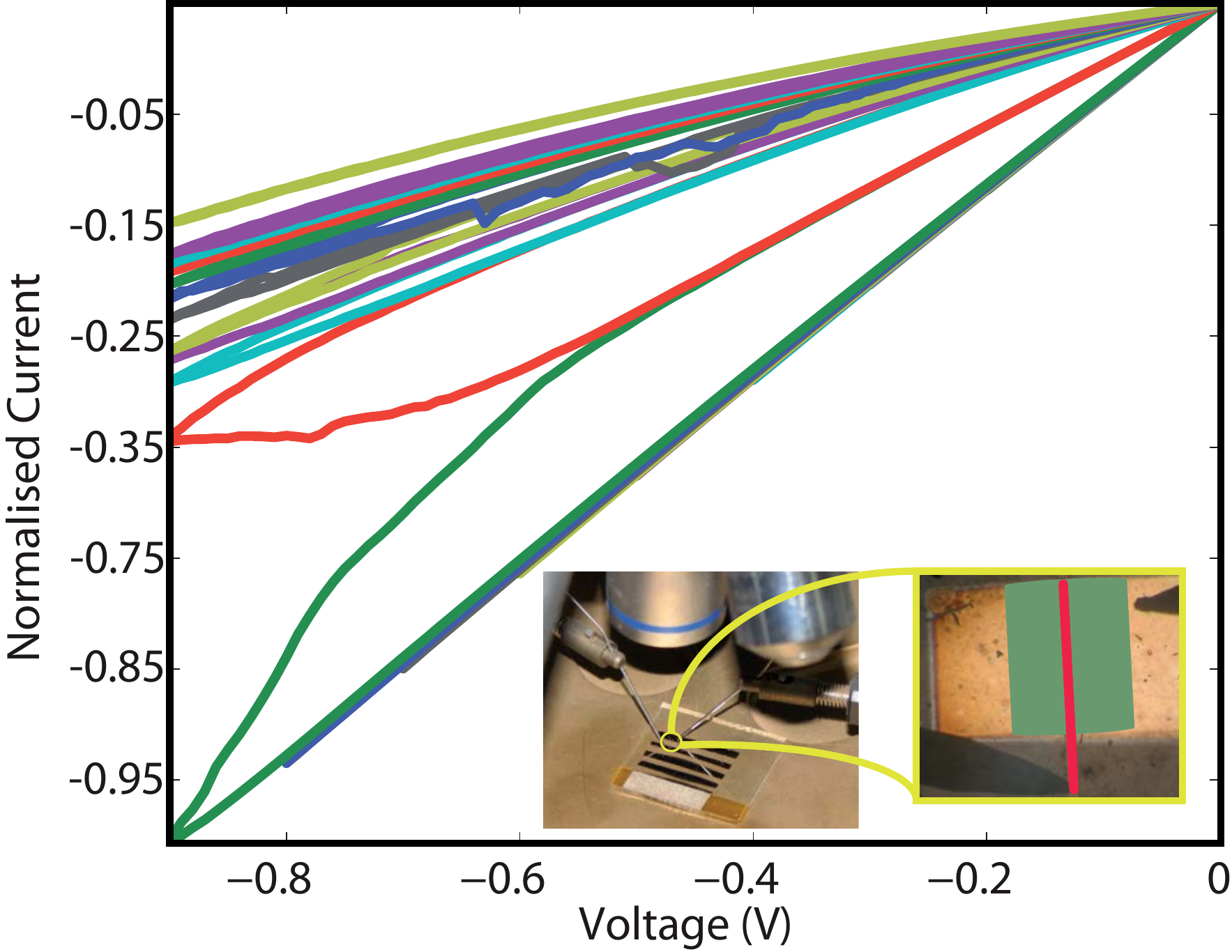}
	\caption{Memristor analogue behaviour. Experimental result from Ag/TiO$_2$/ITO memristor. Current values are normalised to their maximum value ($35~\mu A$). Inset shows a Device Under Test (DUT). The red and the green areas highlight a memristor device.}
 \label{fig:multistable}
\end{figure}

Here we demonstrate such behaviour in Ag/TiO$_2$/ITO experiment, which is an identification for existence of an ionic drift. Fig.~\ref{fig:multistable} illustrates the existence of the multi-stable memory levels. The experiments carried out using a Keithley 4200-SCS. Triangular input voltage was swept from $0~$V to $-0.9~$V and vice versa. Current compliance of $500~\mu$A was applied to avoid any damage to the device. At the end of each cycle device was disconnected from inputs.   

The most critical limitation of analogue memristor is its state decay. Although many stable state can be observed, our measurements for five conductance levels showed decay distribution ranging from a few hours up to a few days. More measurements were not possible with our limited time.  

\subsection{Memristive, plasticity, and learning}

\begin{figure*}[htpb] \centering
 	\includegraphics[width=1.0\textwidth,angle=0]{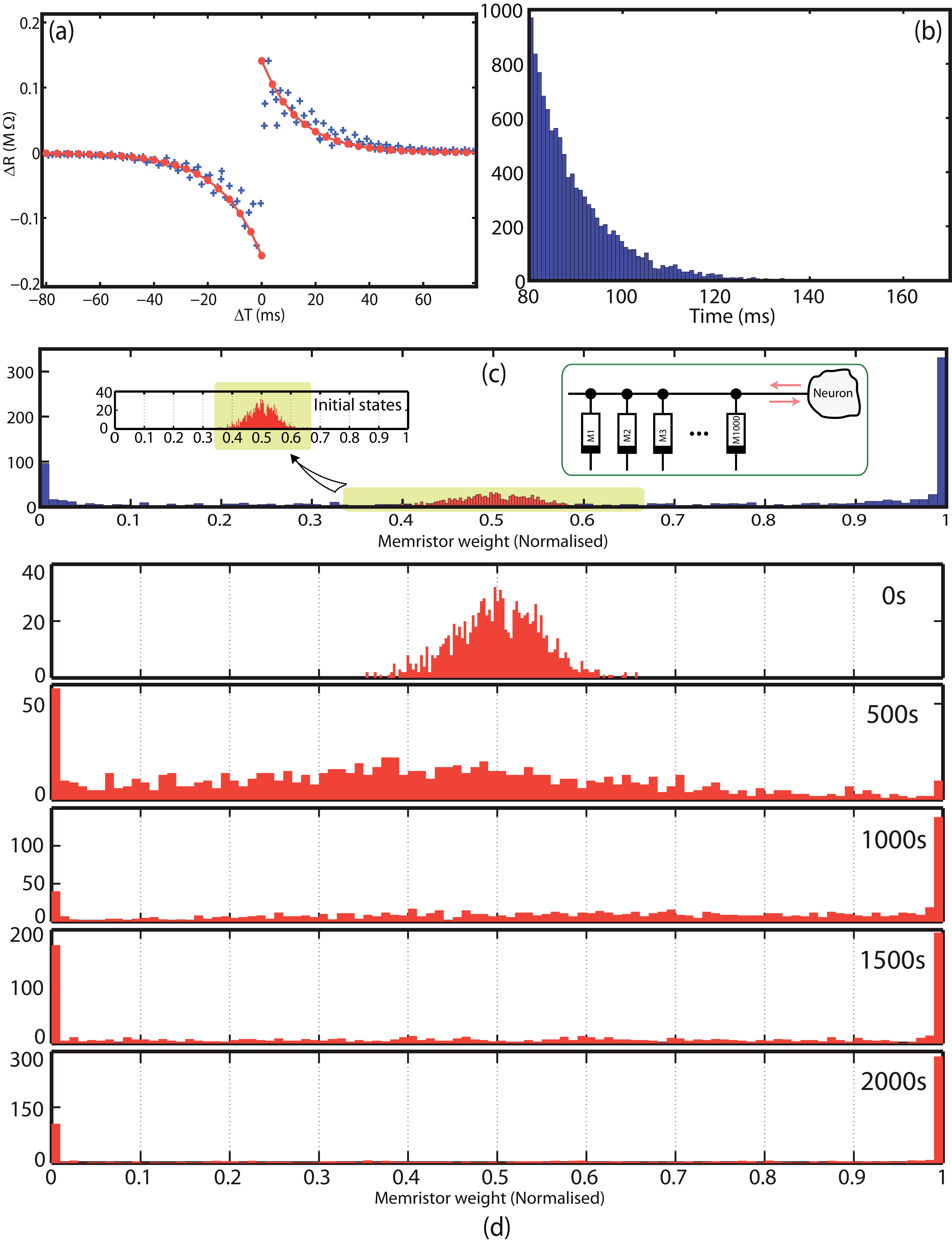}
	\caption{Memristor, plasticity, and competitive learning. (a) Dots illustrate experimental data and the solid (red) line shows the fitting STDP rule. We exclude devices that reach lowest and highest conductances in depression and potentiation processes, respectively, because they add no extra information to our analysis. (b) A Poisson inter-spike interval (ISI) distribution for $1\times 1000$ memristors (synapse) connected to one neuron, inset in (c). (c) Illustrates simulation results of such network. It is clear that it follow the competitive learning behaviour reported in~\cite{competitive:song:2000}. (d) Evolution of synaptic strength from $0$~s to $2000$~s. }
 \label{fig:stdp}
\end{figure*}

The connection can be drawn between memristive devices and biological synaptic update rule, known as STDP, that has been observed in the brain~\cite{nanoscale:jo:2010}. This can be achieved by collecting data from a memristive device based on the time difference, $\Delta t$, between two signals, so called pre- and post-synaptic signals. 
The results are shown in Fig.~\ref{fig:stdp}~(a), which shows how the Device Under Test (DUT) weight (resistance) changes as a function of $\Delta t$. The intermediate states vanish after a certain decay duration whereas a significantly higher potentiation ($x\rightarrow 1$) will be kept as a long term memory. So, the existence of intermediate states decay helps in mimicking the long-term potentiation and short-term plasticity (LTP and LTP) behaviour~\cite{short:ohno:2011}. 

The collected information is then used as stimuli for a network of $1\times 1000$ memristors are connected to one neuron being implemented and pre- and post-synaptic spikes shape is the same as~\cite{spike:zamarreno:2011}, then this network implements the competitive Hebbian learning~\cite{competitive:song:2000}. Initial states have been shown in Fig.~\ref{fig:stdp}~(c) in red. Intentionally, a Gaussian distribution has been employed for the memristors' initial state values. After running the simulation for $35$~minutes, the network results in a population distribution of weights similar to a previously published competitive Hebbian learning rules~\cite{competitive:song:2000}. The additive and multiplicative features of a memristive network strictly depends on the device and its nonlinearity parameters. Fig.~\ref{fig:stdp}~(b) demonstrates a Poissonian ISI distribution.  

\subsection{Programmable analogue circuits}

Although plasticity plays an important role for adaptation and development, networks with fixed synaptic weight pattern should be also studied. Therefore, one of the challenges for this emerging technology is to integrate learning and unlearning hardware as part of a neural computational platform. Since memristors possess a threshold-like behaviour, usually low- or very low-voltage operations do not change the memristor's initial state. This fact helps developing programmable analogue computing circuits~\cite{practical:pershin:2010}. There is also a similar design in~\cite{proposal:mouttet:2009}. The is no simulation or experimental result.

\begin{figure}[htpb] \centering
 	\includegraphics[width=0.8\textwidth]{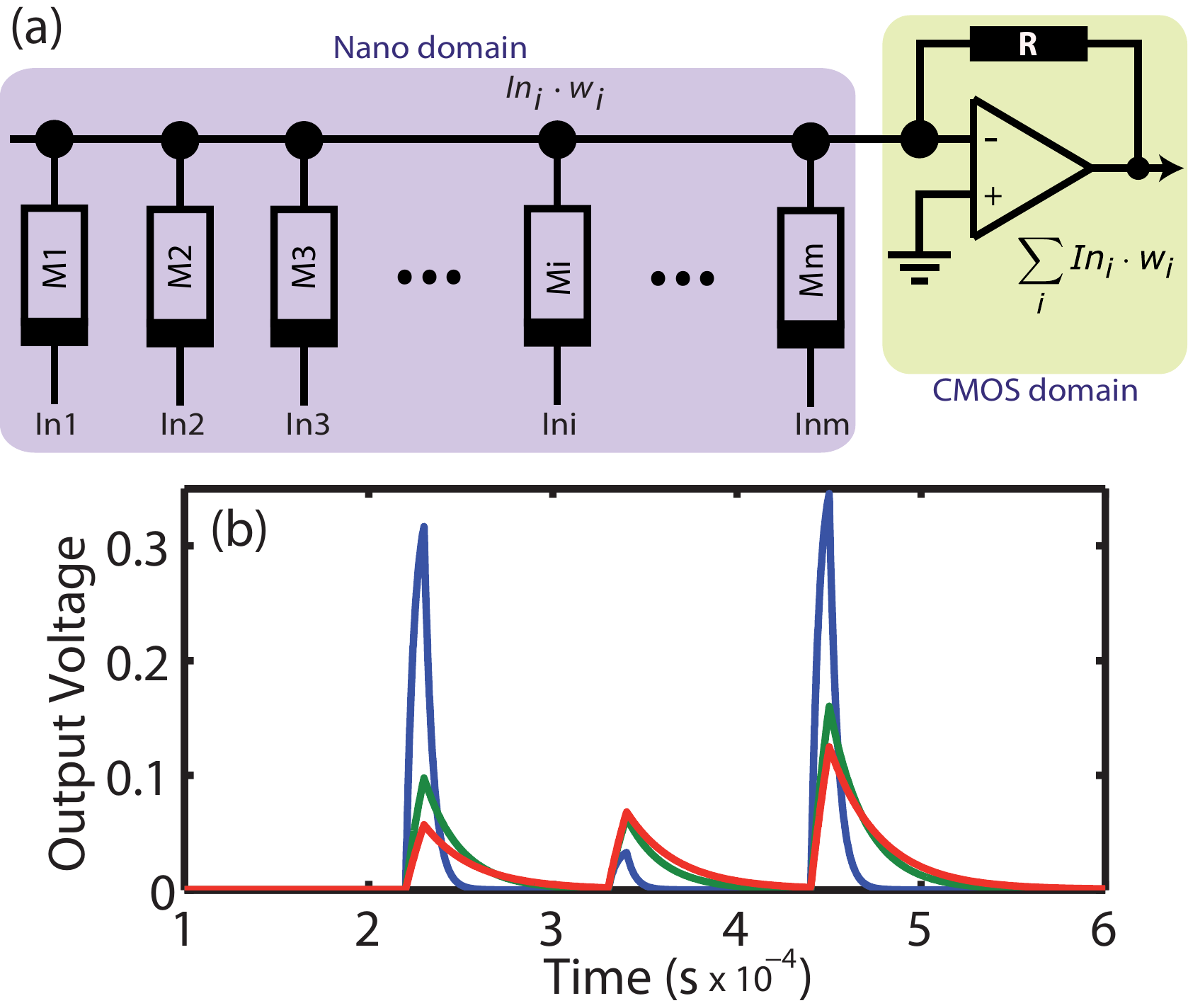}
	\caption{Multiply-accumulate module. (a) Shows a single row of multiply elements (memristors), In$_{i}\cdot w_{i}$. (b) Demonstrates simulation results for two memristors, M1 and M2. In this simulation, memristor M2 programmed at $x=0.5$, which is equivalent to $(R_{\rm HRS}+R_{\rm LRS})/2$. Then memristor M1 changes its resistance from $R_{\rm LRS}$ to $R_{\rm HRS}$ in three steps. Each step is a simulation that is shown with different colours. Blue for $R_{\rm M1}=R_{\rm LRS}$, green for $R_{\rm M1}$ close to $(R_{\rm HRS}+R_{\rm LRS})/2$, and red for $R_{\rm M1}$ close to $R_{\rm HRS}$. The summing amplifier can be replaced by any thresholding module for different applications.}
 \label{fig:proganal}
\end{figure}

Here, we introduce the use of a memristive array for implementing a multiplication of inputs and the memristor's internal state, $w$, which represents the memristor's conductance. Fig.~\ref{fig:proganal}~(a) illustrates a single row of the array and Fig.~\ref{fig:proganal}~(b) shows its simulation results for two elements, M1 and M2, connected to two inputs, In$_{1}$ and In$_{2}$. In this case, we first applied a voltage pulse to M1 to read its conductance, then a pulse to M2 for the same reason. When two voltage pulses are simultaneously applied to M1 and M2, accumulation operation can be clearly observed.    

\subsection{Existence of a threshold-like switching}

\begin{figure}[htpb] \centering
 	\includegraphics[width=1.0\textwidth]{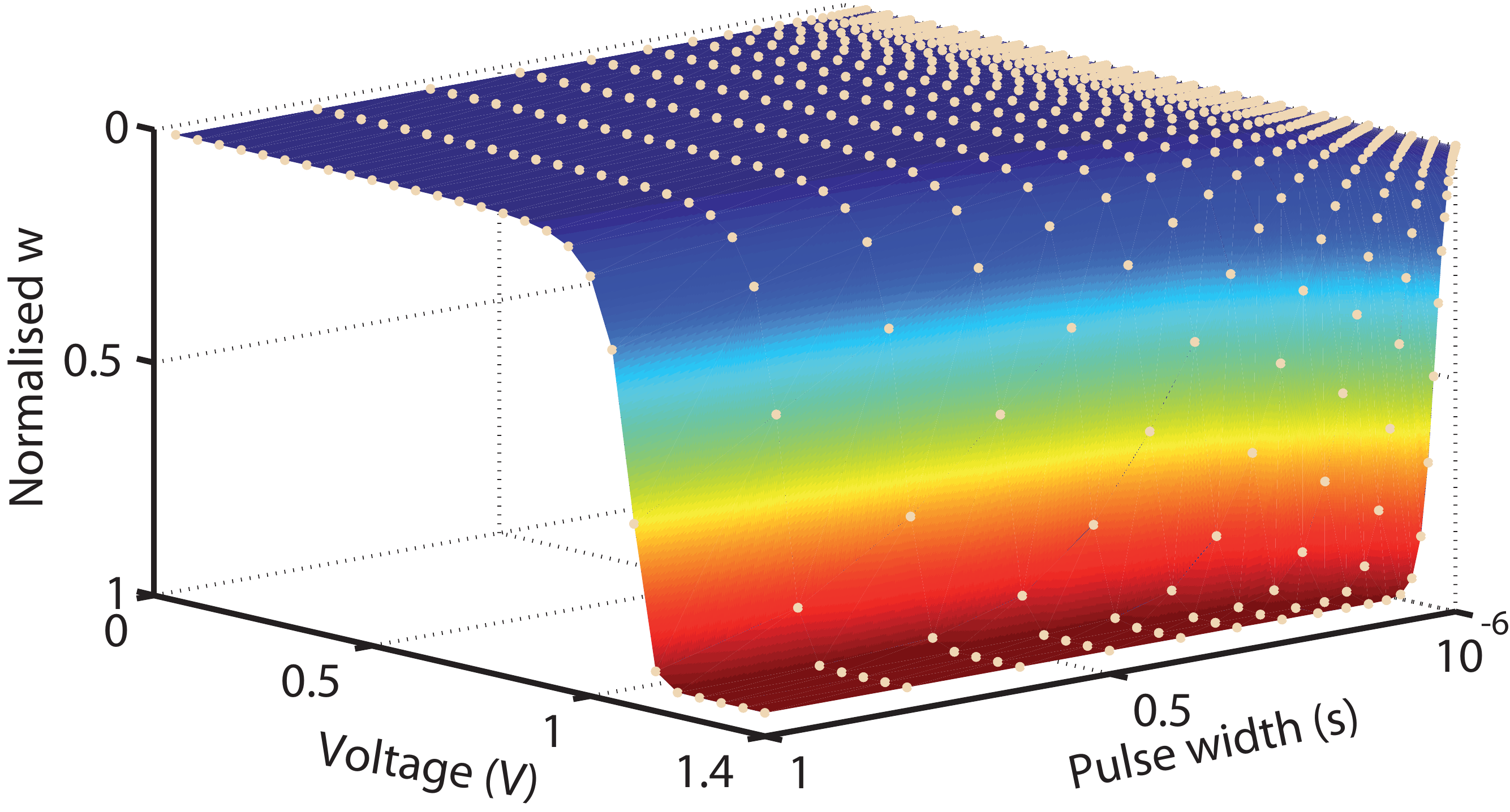}
	\caption{Existance of a switching threshold in the memristor material. The pulse widths are from $10~\mu$s to $1$~s.}
 \label{fig:step3d}
\end{figure}

In this part, we show the existence of a switching threshold in a TiO$_2$-based memristor. According to Chua's definition~\cite{memristor:chua:1971}, memristor links electrical charge to flux, $\varphi$, and $\varphi=\int Vdt$. Therefore, the amount of flux passing through the device can be controlled by $V$ and/or time. So, low pulse widths should not change the conductance if the voltage is lower than a certain value and small voltages similarly do not change the conductance if the applied pulse width is not sufficient. The analysis started from the amorphous (RESET) state and a crystallisation window created above $0.8$~V and $100~\mu$s. Fig.~\ref{fig:step3d} illustrates the results from a Pt/TiO$_2$/Pt memristor. It is observed that the area of crystallisation window decreases as $R_{\rm HRS}$ increases in different devices~\cite{fabrication:kavehei:2011}.      

\section{Digital In-Situ Computing}

The existence of the sharp switching threshold, functional uniformity, intermediate state initialisation, and most importantly state decay creates several problems that can be eliminated or compensated for by using the memristor device as a binary switch.   

\subsection{Complementary Resistive Switch (CRS)}

Although the memristor has introduced new possibilities and it is very well adapted in a crossbar architecture, the inherent interfering current paths between neighbouring cells of an addressed cell impose limitations on the array scalability~\cite{analytical:kavehei:2011,crs:linn:2010}. A possible solution is to build a diode or a transistor in series with a memristor. Using transistors adds other scalability issues due to the fact that transistors are not very well stackable and the application of diodes imposes a high drive current limiting the use of such array in an ultra-low-power applications. 

\begin{figure}[htpb] \centering
 	\includegraphics[width=1.0\textwidth]{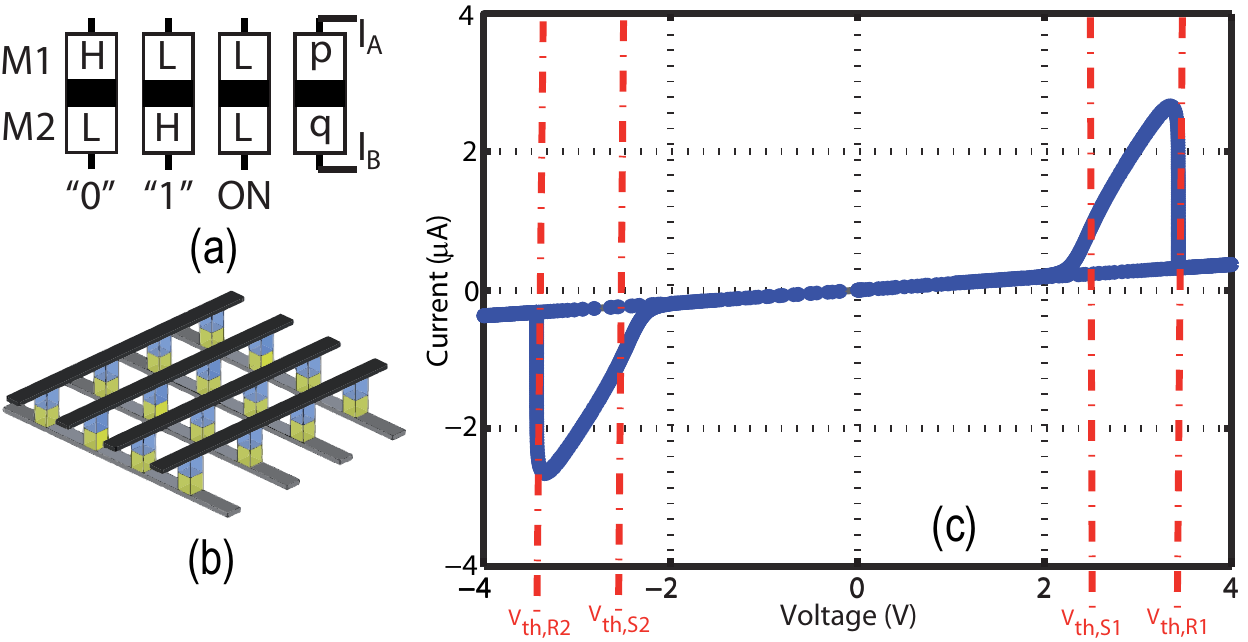}
	\caption{CRS device structure and logical definition of each combination. (a) demonstrates all operational states, (b) illustrates the crossbar view, and (c) shows CRS functionality.}
 \label{fig:crs_states}
\end{figure}

Linn~\emph{et al.}~\cite{crs:linn:2010} introduced a new paradigm by exploiting two anti-serially (with opposite polarities) connected memristors. The structure is similar to a {\it memistor} (note the missing ``r'')~\cite{memistor:widrow:1960,solid:thakoor:1990,memistor:xia:2011}. A (digital) CRS uses a combination of a High Resistance State (HRS) and a Low Resistance State (LRS) to encode logic ``0" and logic ``1". Consequently, the overall resistance of such device is always around HRS, resulting in significant reduction in the parasitic current paths through neighbouring devices. Fig.~\ref{fig:crs_states}~(a) summaries the CRS states. If $p$ and $q$ indicate resistances of the memristors A and B, respectively, four different states can be observed. For example, $p/q\leftarrow$L/H indicates that LRS is written in $p$ (memristor M1) and HRS in $q$ (memristor M2). Combinations L/H and H/L for $p$ and $q$ represent logic ``1'' and logic ``0'', respectively. Note that the H/H state only occurs once in a ``fresh'' device. According to Fig.~\ref{fig:crs_states}~(c) any transition between the states occurs if the applied voltage exceed the SET thresholds, $V_{\rm th,S1}$ or $V_{\rm th,S2}$ and the device's initial state supports the transition. Possible state transitions are shown in Table~\ref{tab:crs_states-trans}, where $p'/q'$ shows the next state, $p/q$ illustrates the initial state, and output is a current pulse or spike. These outputs enable us to have two different read-out mechanisms, logic$\rightarrow$ON or logic$\rightarrow$logic.

The transitions in Table~\ref{tab:crs_states-trans} can be defined using {\it material implication} logic~\cite{memristive:borghetti:2010, crs:rosezin:2011}. It has been proven that implication and FALSE operation are a complete set for logical operations~\cite{principia:whitehead:1912}. This logical operation results in change of $q$ depending on the state of $p$ (or vise versa), known as $p$ IMP $q$, `$p$ implies $q$' or `if $p$ then $q$'. Therefore, if we represent $p$ NIMP $q$ it means `$p$ not implies $q$', Table~\ref{tab:crs_states-trans}~(i), for example, represents $q\leftarrow$H and we say the conditions (initial $p/q$ and $\Delta V$) not implies $q$. 

The destructive read-out should not be a problem for two reasons: (1) refreshing a digital memory is a normal task depends on the decay term and (2) there are no alternative available to combine the CRS properties and a non-destructive read-out.

\begin{table}[htpb] \centering
    \caption{State transitions in a CRS}
    \label{tab:crs_states-trans}

    \begin{small}
    \begin{tabular}{|r|c|c|c|c|}
    \hline \hline
		{} & {$p/q$} & {$\Delta V=V_{{\rm I}_{\rm A}}-V_{{\rm I}_{\rm B}}$} & {$p'/q'$} & {Output} \\ \hline\hline
		{i)} & {``1''} & {$V_{\rm th,S1}<\Delta V<V_{\rm th,R1}$} & {ON} & {pulse} \\ \hline
		{ii)} & {``1''} & {$V_{\rm th,R1}<\Delta V$} & {``0''} & {spike} \\ \hline
		{iii)} & {``0''} & {$V_{\rm th,R2}<\Delta V<V_{\rm th,S2}$} & {ON} & {pulse} \\ \hline
		{iv)} & {``0''} & {$\Delta V<V_{\rm th,R2}$} & {``1''} & {spike} \\ \hline
		{v)} & {ON} & {$V_{\rm th,R1}<\Delta V$} & {``0''} & {--} \\ \hline
		{vi)} & {ON} & {$\Delta V<V_{\rm th,R2}$} & {``1''} & {--} \\ \hline
		\end{tabular}
    \end{small} 
\end{table}

\subsection{CRS-based logical operations}

Here, we introduce CRS-based logical operation and PLA (programmable logic array) that works with the two transitions, logic$\rightarrow$logic and logic$\rightarrow$ON, but we only present it with the later transition. The idea is to charge a bit-line in a crossbar array, and applying inputs to its word lines. The inherent implication property of the device causes a change under certain conditions that we have already discussed. In~\cite{crs:rosezin:2011}, AND and NOR operations are proposed using the logic$\rightarrow$logic transition and current spike read-out process. This method is very dependent on the current spike which occurs by a transient ON state between two logic states. In their implementation, two combinations have been evaluated out of two possible combinations for two CRS devices. Assume voltage, $\Delta V$, is applied across a CRS device that is exceeded its RESET threshold, in this situation this device changes its stored logic, $D$, if $D$ is a certain logic depends on the signature of $\Delta V$. Furthermore, if two CRS devices are connected together, that intermediate point can be connected to either ground or power supply to generate NOR/AND gate. That is the reason that no more possible state can be assumed using such approach.  

\begin{figure}[htpb] \centering
 	\includegraphics[width=1.0\textwidth]{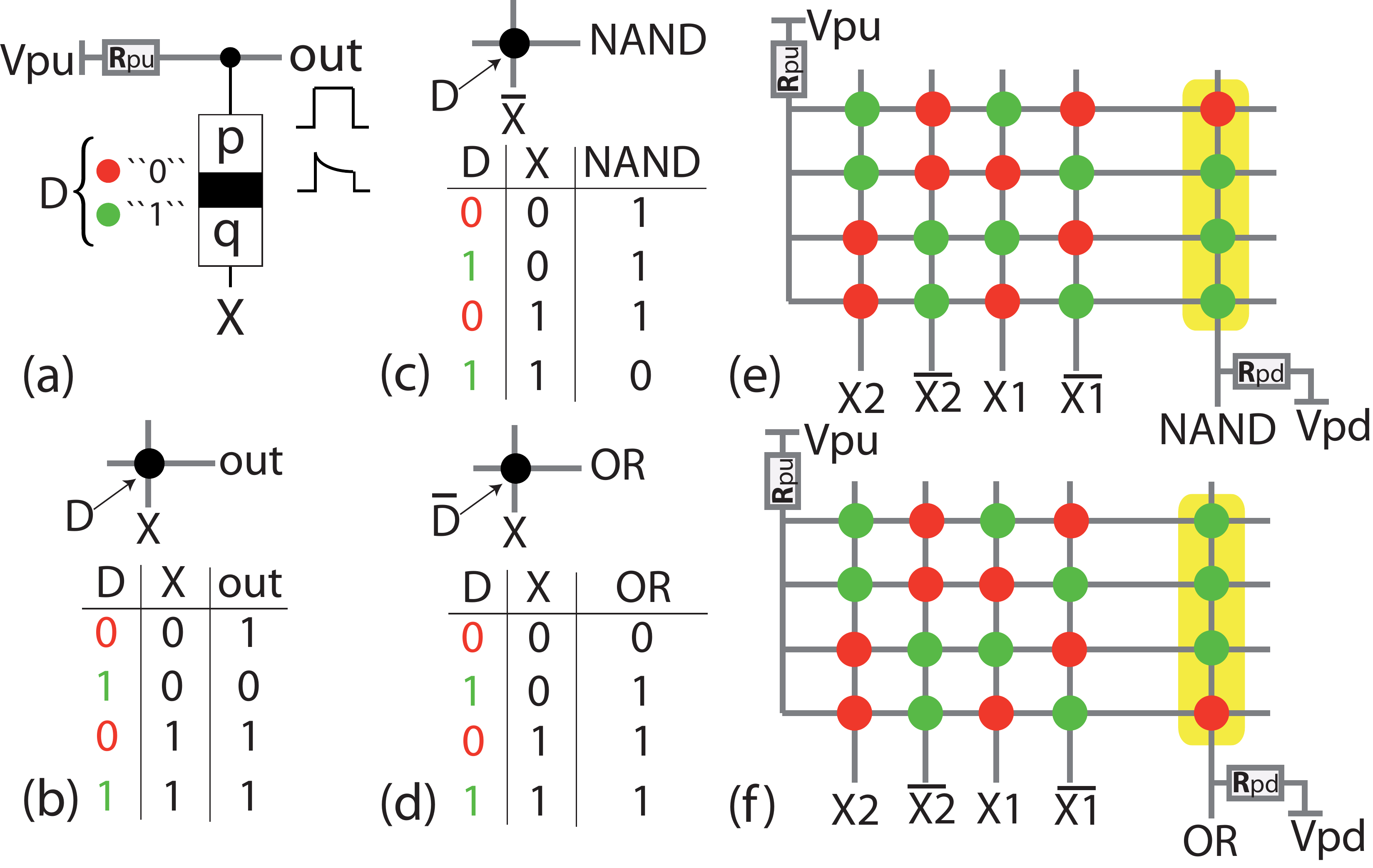}
	\caption{CRS-based logic gate structures. (a) $D$ represents stored data, $X$ is an input, and $R_{\rm pu}$ is pull-up resistor. The output is initially charged and it is discharged depends on $D$ and $In$. (b) Shows how a not implication, NIMP, can be implemented. Here $q'\leftarrow D$~NIMP~$X$. (c) Two inputs NAND gate is implemented by storing one input as device state and another one as an actual input. Here complementary of signal $X$ is applied to the device. (d) Similar to NAND but complementary of $D$ stored in the CRS and $X$ is applied as an input. Therefore, an OR operation implemented, simply by a single CRS device. Obviously, operations are sequential and they requires one (or several) initialisation but this is a drawback for all of the available Boolean logic operations reported in~\cite{crs:rosezin:2011,memristive:borghetti:2010}. Pull-up (charge) voltage is enough to push a device to its ON state and not writing a logic, $V_{\rm th,S}<V_{\rm pu}<V_{\rm th,R}$. NOT function can be also implemented using a single CRS if $D$ stores (the data) $A$ and $X=0$, $F=\overline{A}$. (e) and (d) are PLA implementation of the two logic gates. Here we remove the outputs' complementary signals, AND and NOR. The yellow highlights show the OR-plane and the rest are in the AND-plane.}
 \label{fig:crscomp}
\end{figure}

Here two comprehensive forms of building logical gates are introduced. The first form, allows storing one or more inputs as device state and the second method does not. Fig.~\ref{fig:crscomp} illustrates how CRS works as an implementation of a not implication, NIMP, operation and how NAND and OR operations can be implemented using a single CRS device. Fig.~\ref{fig:crscomp}~(a)-(d) are well explained in the figure's caption and their operations is also described. Fig.~\ref{fig:crscomp}~(e) and (f) follow similar phenomenon but in a form of a PLA. The idea is to have a logic$\rightarrow$ON transition in the OR-plane whenever an output product term is addressed. From the NIMP operation, we know that if the applied inputs are part of the output product terms, that bit-line does not discharged so there will be enough voltage across the output CRS device with stored logic ``1'' (greens) to turn to ON and conduct significantly more current to charge the output signal load. 

In the case of using differential voltage pairs, $V_{\rm pu}=-V_{\rm pd}=1.4$~V was selected as $2.8$~V is the READ voltage (in Fig.~\ref{fig:crs_states}), where $V_{\rm pu}$ and $V_{\rm pd}$ are pull-up and pull-down voltages. Here we applied $V_{\rm pu}=2.8$~V and $V_{\rm pd}=0$~V, so we used $0.25~\mu$m CMOS transistors in our CMOS domain. Therefore, equivalent input voltage for logics ``1'' is $2.8$~V and for logics ``0'' is $0$~V. The pull-up and pull-down resistors, $R_{\rm pu}$ and $R_{\rm pd}$, are both equal to $R_{\rm LRS}\sqrt{2(r+1)}$, where $r=R_{\rm HRS}/R_{LRS}$~\cite{analytical:kavehei:2011}. The used peripheral CMOS circuitries can be found in~\cite{cmos:shakeel:2011}. The sense amplifier was designed for voltage sensitivity more than $100$~mV. 

\begin{table}[htpb] \centering
    \caption{CRS-based logic implementation with two inouts and two CRSs, $F=\overline{D_{1}}\cdot \overline{D_{2}}+\overline{D_{1}}\cdot X_{2}+\overline{D_{2}}\cdot X_{1}+X_{1}\cdot X_{2}$.}
    \label{tab:crslogic}
    \begin{small}
    \begin{tabular}{|r|c|c|c|c|}
    \hline \hline
						 $D_1$ & 					$D_2$ & 					$X_1$ &   				$X_2$ & {Function} \\ \hline\hline
	  $\overline{A}$ & 						$A$ &  						$0$	& 				 		$B$	& $A\cdot B$ (AND) \\ \hline
							 $A$ & $\overline{A}$ &  						$0$	&  $\overline{B}$	& $\overline{A+B}$ (NOR) \\ \hline
							 $A$ & $\overline{A}$ &  $\overline{B}$	& 					${B}$	& ${A\oplus B}$ (XOR) \\ \hline
		$\overline{A}$ & 						$A$ &  $\overline{B}$	& 					${B}$	& ${A\odot B}$ (XNOR) \\ \hline
		\end{tabular}
    \end{small} 
\end{table}

\begin{figure}[htpb] \centering
 	\includegraphics[width=1.0\textwidth]{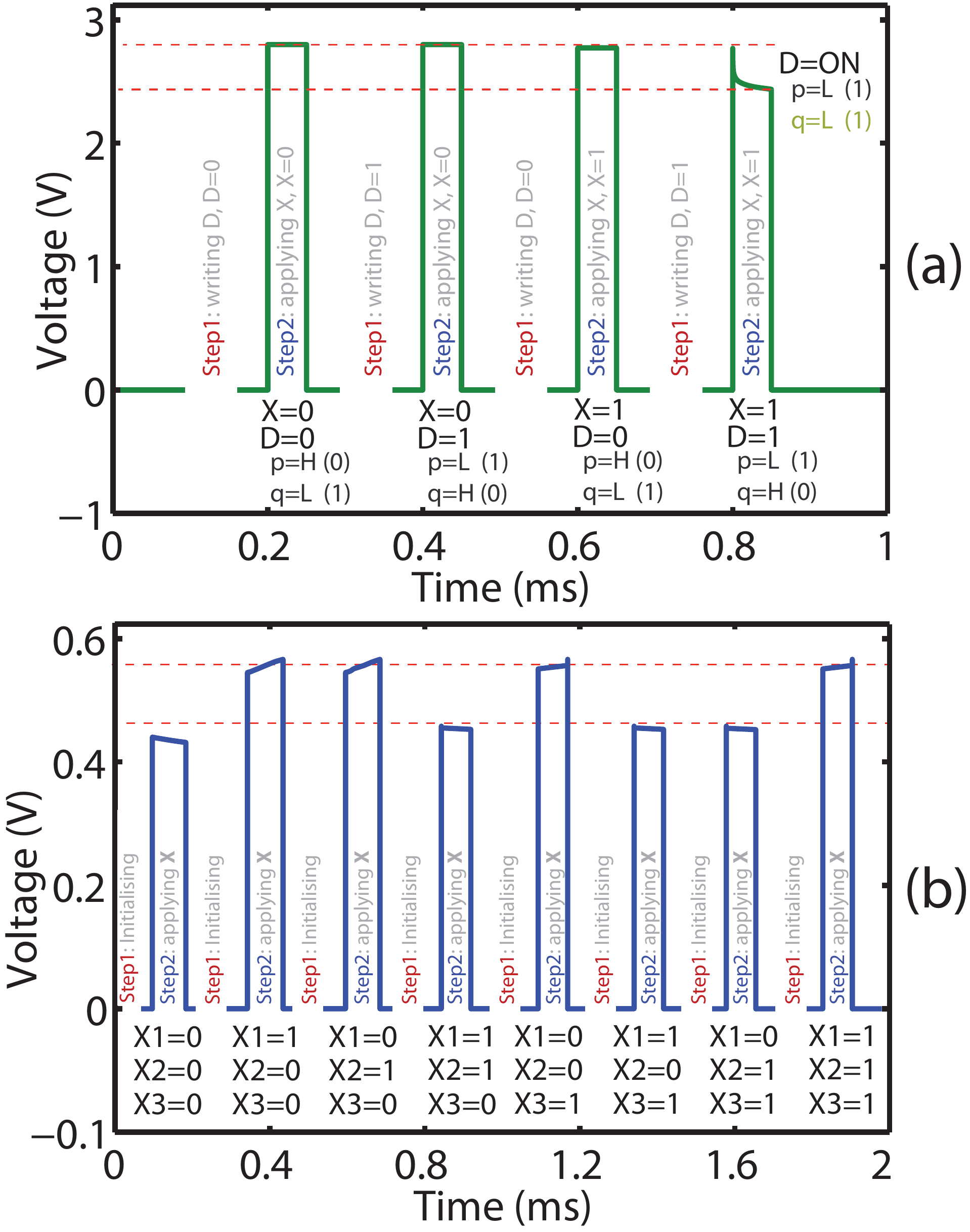}
	\caption{CRS-based logic gate simulations. (a) A 2-input NAND gate (Fig.~\ref{fig:crscomp}~(c)) simulation. In this style, we are allowed to store one input as the CRS state. (b) A 3-input XOR (SUM) function, implemented in a PLA structure. In both cases, (a) and (b), dashed red line show worst-case low and high output voltages that are sent to sense amplifiers. Due to limited space, complementary output, XNOR, is not shown. Initialisation in (b) means, the array should be initialised before the next logical operation and this is the main reason that the first approach (in (a)) is a far more efficient implementation in terms of both hardware and number of steps. No initialisation is required in (a), because 'writing $D$' effectively means writing one of the input's logic into the device.}
 \label{fig:digitalcomp}
\end{figure}

Assuming we have two inputs, $X_{1}$ and $X_{2}$, and two CRS devices, $D_{1}$ and $D_{2}$, connected to these inputs and a charged bit-line. A number of functions can be implemented by writing $\overline{F}=D_{1}\cdot \overline{X_{1}}+D_{2}\cdot \overline{X_{2}}$, hence, $F=\overline{D_{1}}\cdot \overline{D_{2}}+\overline{D_{1}}\cdot X_{2}+\overline{D_{2}}\cdot X_{1}+X_{1}\cdot X_{2}$. The first term, $\overline{D_{1}}\overline{D_{2}}$, indicates that if both CRSs store ``0'' TRUE ($F=1$) is implemented. Some other function that is implemented using this configuration are shown in Table~\ref{tab:crslogic}. In~\cite{high:kavehei:2011} we demonstrates a CRS-based content addressable memory based on the XOR/XNOR function. Fig.~\ref{fig:digitalcomp}~(a) illustrates simulation of a two input NAND function. The most significant advantage of this method is that the initialisation step (step 1) which is writing data into CRS arrays and not a simple refreshing cycle. While in a PLA structure, Fig.~\ref{fig:digitalcomp}~(b), the initialisation is a refreshing cycle. Furthermore, in computer arithmetic operations signals arrive with relative delays, like SUM results and CARRY output, that can be used in parallel with the programming of CRS arrays.  

\section{Conclusion}

This paper introduced basic functional blocks for analogue and digital computation based on memristive devices. It is difficult to have a fair comparison between emerging and the conventional devices as the emerging technologies are at their early stages. Moreover, architectural aspects for future computers seems to be dependent to the concept of universal memory and computational capability of one individual device or nano-system that is entirely different with the classical von Neumann computational framework. Therefore, introducing more compatible circuits and algorithms with these futuristic technologies could play an important role. 
This work presented the existence of ionic drift in the fabricated memristors. We have also illustrated how the memristor can be used to implement competitive Hebbian learning (additive STDP). An analogue multiply-accumulation circuit was introduced that is able to implement a low precision multiplication and addition. This circuit combines inherent non-volatile memory and dynamics of a memristor as a synapse. The problem of state decay then results in developing a digital version of such learning system which is out of the scope for this paper. However, the idea of digital computing using a more robust memristive device, CRS, was explained and two methods for implementing logical blocks were introduced.  

\section*{Acknowledgement}

This work was supported by Australian Research Council (ARC) and grant No.R 33-2008-000-1040-0 from the World Class University project of MEST and KOSEF through Chungbuk National University.

\end{document}